\documentclass[twocolumn,prc,showpacs,amsmath,amssymb,nofootinbib,floatfix]{revtex4}

\usepackage{graphicx}
\usepackage{dcolumn}
\usepackage{bm}

\newcommand{\smfrac}[2]{\mbox{$\textstyle\frac{#1}{#2}$}}
  
\newcommand{\bra}[1]{ \mbox{$\langle#1|$} }

\newcommand{\tp}{\mbox{$2^+$} }

\begin{document}

\title{The Skyrme energy functional and low lying $\mathbf{2^+}$ states in
       Sn, Cd and Te isotopes}

\author{P. Fleischer}
\author{P. Kl\"upfel}
\author{P.-G. Reinhard}%
 \email{reinhard@theorie2.physik.uni-erlangen.de}
\affiliation{%
Institut f\"ur Theoretische Physik II, Universit\"at Erlangen,
D-91058 Erlangen, Germany\\
}%
\author{J. A. Maruhn}
\affiliation{%
Institut f\"ur Theoretische Physik, Universit\"at Frankfurt,
D-60325 Frankfurt, Germany\\
}%

\date{\today}

\begin{abstract}
We study the predictive power of Skyrme forces with respect to low
lying quadrupole spectra along the chains of Sn, Cd, and Te isotopes.
Excitation energies and $B(E2)$ values for the lowest quadrupole
states are computed from a collective Schr\"odinger equation which as
deduced through collective path generated by constraint
Skyrme-Hartree-Fock (SHF) plus self-consistent cranking for the
dynamical response. We compare the results from four different Skyrme
forces, all treated with two different pairing forces (volume versus
density-dependent pairing).
The region around the neutron shell closure $N=82$ is very
sensitive to changes in the Skyrme while the mid-shell isotopes
in the region $N<82$ depend mainly on the adjustment of pairing.
The neutron rich isotopes are most sensitive and depend on both
aspects.
\end{abstract}

\maketitle

%

\section{introduction}

A key feature of nuclear excitations are the low-lying $2^+$ states.
Their properties delivered crucial input for developing an
understanding of nuclear structure \cite{Eis70aB,BM2B}. At first
glance, they suggest the collective picture of the nucleus as a liquid
drop which can undergo global quadrupole oscillations and which freeze
under certain conditions to a stable rotator. This view has been
formulated in terms of the Bohr-Hamiltonian which establishes a
collective dynamics in the five quadrupole degrees of freedom
\cite{Boh52}. The parameters of the collective Hamiltonian have to be
adjusted phenomenologically, see e.g. the applications in
\cite{Gne71}. The collective approach has been revived with the
interacting boson model (IBM) which has found widespread application
and proven to be extremely useful in sorting nuclear low-energy
spectra \cite{Iac87aB}.

The collective picture is seemingly in contrast to the microscopic
view which sees the nucleus as consisting out of shells of single
nucleons arranging themselves in a common mean field
\cite{Goe48,Hax49}. The views can be unified by the concept of a
deformed mean field which establishes a relation between
single-particle shell structure and global deformations
\cite{Bro71aB,Row70aB}. The collective motion is then understood as
vibration (or rotation) of the mean field similar to the
Born-Oppenheimer method for describing molecular vibrations. The
connection is established on a formally sound level by the generator
coordinate method (GCM) \cite{Hil53a,Gri57a} which describes
collective dynamics as coherent superposition of a continuous set of
deformed mean-field states, called the collective path. The GCM within
the Gaussian-Overlap-Approximation (GOA) allows to establish contact
between the microscopic foundation and a collective Bohr-Hamiltonian
\cite{Rei87aR,Bon90b}.  Starting from the GCM, the lines of
applications spread enormously. There are, on the one hand, fully
fledged GCM calculations which skip the collective Hamiltonian as
intermediate level and compute low-energy spectra directly from the
coherent superposition of the collective path; these sophisticated
calculations imply exact projection for the conserved quantities, as
particle number, angular momentum, and center of mass; there are many
published results around, we mention here \cite{Val00a,Rod02a} as two
recent examples. On the other hand, one finds several approximations
to the microscopic computation of the Bohr-Hamiltonian; most
applications hitherto employ a phenomenological shell model to
describe the deformed mean field, see e.g. \cite{pomorski}. There are
also several self-consistent calculations along that line, for an
early example see \cite{Gir82a} and for more recent achievements
\cite{Liber99,Pro04a}.

An alternative direction of development remains at the microscopic
mean-field description and makes it manageable by restricting
considerations to small amplitude motion. This yields the much
celebrated random-phase approximation (RPA) which has its stronghold
in the description of giant resonances, see e.g. \cite{Rin80aB}.  The
appropriate extension to non-closed shells with pairing is the
quasi-particle RPA (QRPA) which has only recently been developed up to
a rigorously self-consistent level \cite{Ter04a}. The QRPA describes
formally the whole excitation spectrum, including the low lying $2^+$
states, and it optimizes all states automatically. It assumes, however,
small amplitudes, i.e. harmonic motion. This is perhaps legitimate
close to magic nuclei but somewhat dubious elsewhere. The above
mentioned theories for large amplitude collective motion concentrate
on the lowest state only but try to take into account all effects of
anharmonicity due to soft potential energy landscapes and shape
isomerism. The fully fledged adiabatic time-dependent Hartree-Fock
method (ATDHF) provides an unambiguous optimization scheme for the
large-amplitude collective path \cite{Goe80b,Rei87aR}.  However, that
rather involved scheme has not yet been used for heavy nuclei as we
are going to study here.  We use presently ATDHF only to compute the
self-consistent collective mass and employ the more intuitive
constraint Hartree-Fock to generate the path.

The connection from a microscopic Hamiltonian to collective spectra 
via a large-amplitude collective path is
well established by virtue of the GCM. An open problem is the
microscopic input. Self-consistent nuclear mean-field models employ
effective energy functionals as e.g. the Skyrme-Hartree-Fock method,
the Gogny force, or relativistic mean field, for a recent review see
\cite{Ben03aR}. These are empirically adjusted to nuclear ground state
properties of stable nuclei. There exists a large manyfold of
equivalent parameterizations which provide comparable ground state
properties but can differ substantially in predictions to exotic
nuclei or resonance excitations \cite{Rei02ar,Ben03aR}. It is by no
means guaranteed that all mean-field parameterization produce at once
the correct collective low-energy vibrations. The contrary is to be
expected, namely a broad span of predictions amongst which only a few
parameterizations deliver a satisfying spectrum. To phrase that
positively: low-energy vibrations provide useful information for a
better selection of mean-field parameterizations. We aim here at a
first exploration of the connection between mean-field
parameterizations and emerging low-energy spectra. We do that for the
Skyrme-Hartree-Fock approximation by comparing the results of several
different Skyrme forces and pairing recipes.

It is obvious that such systematic studies need to confine the subject
and the method in order to keep things manageable. As test cases, we
consider the lowest 2$^+$ state in the chain of Sn isotopes and its
even neighbors Cd and Te. These share basically one type of collective
motion being predominantly soft vibrators. For the practical
technique, we employ GCM-GOA through a microscopically computed
Bohr-Hamiltonian. For reasons of simplicity, the microscopic
information is computed along axially symmetric shapes and
interpolated into the full space of quadrupole degrees of freedom.
This approximation allows large scale scans and it is acceptable for
soft vibrators as they are studied here.

\section{Formal framework}\label{sec:formalframe}

\subsection{Underlying microscopic model - input parameters}

As starting point, we take a microscopic mean-field theory at the
level of the Skyrme-Hartree-Fock model augmented by pairing in BCS
approximation plus Lipkin-Nogami correction for approximate particle
number projection.  This is a standard approach in nuclear structure
physics. We refer the reader to \cite{Ben03aR} for a detailed
description of the energy functional and subsequent mean-field
equations. We recapitulate here only briefly the spectrum of variants
of that model which will play a major role in the following
discussions.

The mean-field part is determined by the Skyrme energy functional
$E_{\rm Sk}(\rho_\nu,\tau_\nu,{\cal J}_\nu,j_\nu,\sigma_\nu)$ which
depends on the local density $\rho_\nu$, kinetic-energy density
$\tau_\nu$, spin-orbit density ${\cal J}_\nu$, current $j_\nu$ and
spin-density $\sigma_\nu$ and where $q$ means protons or neutrons. The
functional form is basically settled since two decades \cite{Bar82a}
with minor extensions in later stages
(e.g. \cite{Rei95a,ChabanatDiss}).
However, there exists a great variety of actual parameterizations for
the Skyrme energy functional.  Most of them provide a high-quality
description of nuclear bulk properties as binding energies and
radii. They differ in details as, e.g., isovector forces or surface
properties. We are going here to apply the Skyrme functionals to a
regime far from what had been considered in the fits. It is thus
important to explore a minimal variation of parameterizations within
the Skyrme framework. We will consider here: SkM$^*$ as a widely used
traditional standard \cite{Bar82a}, Sly6 as a recent fit which
includes information on isotopic trends and neutron matter
\cite{Cha97a}, SkI3 as a fit which maps the relativistic isovector
structure of the spin-orbit force and takes care of the surface
thickness \cite{Rei95a}, and SkO \cite{Rei99b} as a recent fit relying
an the same fit data as SkI3 but with additional constraint on the
two-nucleon separation energies around $^{208}$Pb and with a better
adjusted asymmetry energy.
That selection contains a large span of effective masses: SkI3
$\leftrightarrow m^*/m=0.6$, SLy6 $\leftrightarrow m^*/m=0.7$, SkM$^*$
$\leftrightarrow m^*/m=0.8$, and SkO $\leftrightarrow m^*/m=0.9$.  The
effective mass has an influence on the level density near the Fermi
surface which, in turn, may have an effect on the low-energy
collective states.  There is also a difference in the isovector and
spin-orbit properties.  Besides the effective mass and asymmetry, the
bulk parameters (equilibrium energy and density, as well as
incompressibility) are comparable.

The second key ingredient is pairing. A present-day standard is to use a zero
range pairing force often called volume
pairing. We will use the notion $\delta$-interaction (DI) pairing. 
A widely used variant for the pairing force 
is a density-dependent delta-interaction (DDDI) \cite{Ter95a}.
Both recipes are summarized as
\begin{equation}
 V^{\rm(pair)}= 
 \left\{\begin{array}{l}
   V^{\rm(DI)}_\nu\delta(r_1-r_2)
 \\
 V^{\rm(DDDI)}_\nu\delta(r_1-r_2)\left[1-\rho(\bar{r})/\rho_{0}\right]
 \end{array}
 \right.
\end{equation}
The pairing strengths  or $V^{\rm(DDDI)}_\nu$ are
adjusted to odd-even staggering of binding energies in a few
representative semi-magic nuclei (Sn and Pb isotopes, N=82
isotones). The adjustment is done for each force separately because
the much different effective masses call for different pairing
strengths. 
The actual values used here are given in table \ref{tab:pairstr}.
\begin{table}
 \begin{tabular}{|l|rr|rr|}
\hline
 & $V^{\rm(DI)}_p$ & $V^{\rm(DI)}_n$ & $V^{\rm(DDDI)}_p$ & $V^{\rm(DDDI)}_n$\\
\hline
 SkM$^*$ & 279.1 & 259.0 &  990.0 &  802.0 \\
 SLy6 &  298.8 &  288.5 &   1053.1 &   864.2 \\
 SkI3 &  335.4 &   331.6 & 1233.0 & 996.0 \\
 SkO  &  253.0 & 269.0 & 1007.4 & 893.7 \\
\hline
 \end{tabular}
\caption{\label{tab:pairstr}
Pairing strengths for the two pairing recipes and for the Skyrme
forces used in this paper. The strengths are given in units of
fm${-3}$.
}
\end{table}
The pairing recipe is to be augmented by a cutoff in single particle
space. We use a smooth cutoff with a Woods-Saxon profile in the single
particle energies. The switching energy is chosen such that the
pairing space covers 1.6$N^{2/3}$ particles above the Fermi energy,
for details see \cite{Ben00c}.  In order to explore the influence of
the pairing recipe, we will also discuss deliberate rescaling of the
pairing strengths.

\subsection{Deduced collective dynamics}

The mapping from the microscopic to a collective description is
performed with the generator-coordinate method (GCM). This is a much
celebrated method in nuclear structure physics, for a review see
e.g. \cite{Rei87aR} and for a brief summary \cite{Ben03aR}. We outline
here the basic steps and provide a more detailed compact account in
appendix \ref{app:method}.

The stationary mean field equations as such provide only a few well isolated
states, preferably the ground state and perhaps some isomers.  Each state is
characterized by one BCS wavefunction $|\Phi\rangle$ which is composed of a
set of single-nucleon wavefunctions together with their occupation amplitudes.
In order to describe motion, one needs to consider a time-dependent mean-field
theory, in the nuclear community often called time-dependent Hartree-Fock
(TDHF). Large amplitude collective motion is related to low energy
excitations, thus slow motion. This justifies the adiabatic limit known as
It yields at the end, a collective path $\{|\Phi_q\rangle\}$
where $q$ stands for continuous series of deformations, predominantly of
quadrupole type because nuclei are softest in that degree of freedom.  The
dynamical aspect is added in first order of collective velocity, i.e. in terms
of linear response to a collective displacement. It is a widely used
approximation to determine the collective path from quadrupole constrained
Hartree-Fock-BCS (CHF). 

The link between microscopic description and collective dynamics is
established by the collective path $\{|\Phi_q\rangle\}$ where $q$
stands for a continuous series of deformations, predominantly of
quadrupole type because nuclei are softest in that degree of freedom.
A systematic theory for an optimized collective path is provided by
adiabatic TDHF (ATDHF) \cite{Bar78a,Goe78a,Kle91a}. The overwhelming
majority of practical applications simplifies the construction by
using a simple quadrupole constraint Hartree-Fock to produce the
$|\Phi_q\rangle$.  The path, once established, serves as a basis along
which the collective motion expands. The corresponding microscopic
state is described as a collective superposition
$|\Psi\rangle=\int dq\,|\Phi_q\rangle f(q)$.
The state $|\Psi\rangle$ is optimized by a variational principle.
This is the fully fledged GCM which can be those days attacked in a
straightforward numerical manner, see
e.g. \cite{Bon91a,Val00a,Rod02a}.  However, that is still a very
demanding task and not so well suited for broad surveys as we intend
it here.  We use here as a simple, efficient, and reliable shortcut
the Gaussian overlap approximation (GOA) which parameterizes the norm
and Hamiltonian overlap in terms of Gaussians, e.g. for the norm
overlap as
$\langle\Phi_q|\Phi_{q'}\rangle=\exp{\left(-\lambda(q-q')^2/4\right)}$.
It provides an acceptable approximation, particularly for medium and
heavy nuclei \cite{Rei87aR,Hag02}. The GCM-GOA yields at the end a
fairly simple collective Hamiltonian where the collective potentials
and masses are unambiguously computed from the microscopic energy
functional and the collective path.  Quadrupole motion has five
degrees-of-freedom \cite{Boh52,Gne71,Iac87aB}.  The emerging
collective Hamiltonian thus has the form of a generalized
Bohr-Hamiltonian while its potentials and masses are computed from
microscopic input \cite{pomorski,Liber99,Pro04a}.

The practice of GCM-GOA is a bit involved, see the appendix for a few
more details. We summarize here the steps: The energy expectation
value along the path yields a raw collective potential ${\cal
V}(q)$. The collective mass and moments of inertia are obtained by
dynamical linear response about a given point at the path often called
self-consistent (or ATDHF) cranking \cite{Goe78a}; as an approximation,
Inglis cranking is used whenever justified. Zero-point energy
corrections to the potential are computed from these masses and the
collective fluctuations (quadrupole, angular momentum) of the states
$|\Phi_q\rangle$.
In fact, a topologically corrected GOA is used to allow a numerical
robust computation of potentials and masses in the intrinsic frame
(defined by a diagonal inertia tensor) \cite{Rei78b,Hag02}. This
provides an interpolation scheme to connect safely the near spherical
shapes with larger deformations.

After all, we restrict the microscopic calculations to axial
symmetry. The fully five-dimensional quadrupole dynamics is recovered
by interpolation of the collective potential and masses between
prolate and oblate shapes into the triaxial plane. This approximation
saves two order of magnitude computation time and thus allows the
large scale systematics as we intend it here. On the other hand, it is
well justified at and in the vicinity of the spherical shape. The test
cases for the present study are Sn isotopes and its even-even
neighbors Cd and Te which are predominantly soft vibrators around
spherical mean. Moreover, we confine the study to the first excited
$2^+$ state (and occasionally to the $0^+$ ground state) which both are
not very sensitive to details in the triaxial plane. All that
considered, the triaxial interpolation is a useful and legitimate
approximation for the intended systematic explorations.

\subsection{An example for potentials and masses}

\begin{figure}
\centerline{\includegraphics[width=8cm]{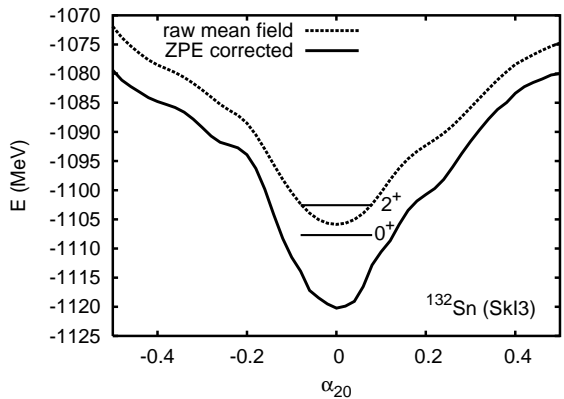}}
\caption{\label{fig:pots}
The raw collective potential ${\cal V}$ and the effective
potential $V$  including the
zero-point energy corrections (\ref{eq:rotvibZPE}), both
drawn as function of the intrinsic axial quadrupole momentum
$a_{20}$. Test case is  $^{132}$Sn  computed with
DI pairing and SkI3.
The position of
the $0^+_0$ ground state and the first $2^+$ state are indicated by
horizontal bars. The difference between the minimum of ${\cal V}$ and
the $0^+_0$ energy is the correlation energy $\Delta E_{\rm corr}$.
}
\end{figure}
In order to exemplify details of the calculations, figure
\ref{fig:pots} shows for the case of $^{132}$Sn the collective
potential before and after zero-point energy correction (ZPE)
(\ref{eq:rotvibZPE}). The ZPE induce obviously a strong global
down-shift in energy because the spurious energy content from
collective fluctuations in the $|\Phi_q\rangle$ is
subtracted. Moreover, they may change the shape of the potential. The
corrected potential has its minimum at a slightly deformed position
although the doubly magic $^{132}$Sn is a perfectly spherical nucleus
in a pure mean-field description (see the well defined spherical
minimum in ${\cal V}$). This is the same effect as happens in
variation after rotational projection (for a model discussion see
\cite{Hag02}): knowing that the projection restores spherical shape
anyway, the system takes advantage of a small deformation to acquire
correlation energy. It is comforting that we see the same effect here
because our treatment of quantum correction should include a good
approximation to rotational projection.

Figure \ref{fig:pots} also indicates the position of the $0^+_0$
ground state and the first excited $2^+$ state. The $0^+_0$ lies above
the bottom of the intrinsic potential $V$ as it should be, to account
for the correct physical zero-point energies, but it stays below the
minimum of the raw potential ${\cal V}$ because the larger spurious
zero-point energy had been subtracted before adding the physical one.
The net effect is a correlation energy $\Delta E_{\rm corr}$ which
expresses that the collectively correlated ground state is better
bound than the mean-field ground state. The $2^+$ state lies, of
cause, above the $0^+_0$ state. The quantity of interest here is the
excitation energy $E(2^+)$ which is computed as the difference between
the total $2^+$ energy and the $0^+_0$ energy.

\section{Results}

\subsection{Results for the chain of Sn isotopes}
\label{sec:e2_sn}

\subsubsection{Variation of forces}

\begin{figure}
\unitlength1mm
\centerline{\includegraphics[width=85mm]{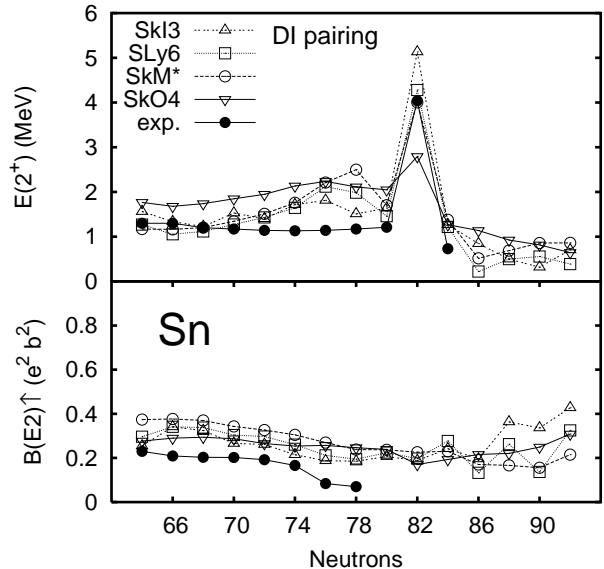}}
\caption[zu]{\label{pic:all-e2p}
  Energies $E(2^+)$ and $B(E2)\!\!\uparrow$ values (= $|
  \langle0^+|Q_{2M}|2^+\rangle |^2$) along the chain of Sn isotopes
  calculated using the four different Skyrme interactions as
  indicated.  The experimental results are taken from \cite{nudat}.  }
\end{figure}
Figure \ref{pic:all-e2p} shows the $2^+$ excitation energies and
transition strengths along the chain of Sn isotopes for the four
chosen SHF parameterizations and in comparison to experimental
data. At first glance, we see that all calculations hit the right
order of magnitude. They also reproduce the increase of the $E_{2^+}$
at the shell closure $N=82$. At closer inspection, however, we see
interesting differences and mismatches in detail.

\begin{table}
\begin{tabular}{|l|llll|}
\hline
 force & SkI3 & SLy6 & SkM$^*$ & SkO \\
\hline
 $E(2^+)$ [MeV] &  4.36 & 3.76 &  3.94 & 2.41 \\
 spectral gap protons [MeV] & 6.6 & 6.2 & 6.4 & 6.0 \\
 spectral gap neutrons [MeV] & 6.6 & 6.0 & 5.4 & 4.0 \\
 $m^*/m$ & 0.6 & 0.7 & 0.8 & 0.9 \\
\hline
\end{tabular}
\caption{\label{tab:e2p}
Comparison of the $E(2^+)$ energies in $^{132}$Sn for the various
Skyrme forces with the spectral gaps  in $^{132}$Sn and the effective
mass $m^*/m$ associated with the forces.
}
\end{table}
Let us first concentrate on the doubly magic case of $N=82$.  Shell
effects directly related to the SHF forces should dominate here.  And
indeed, we see that the $E(2^+)$ are closely related to spectral
properties. Table \ref{tab:e2p} shows the $E(2^+)$ energies in
comparison to the spectral gaps of protons and neutrons in
$^{132}$Sn. The spectral gaps are the energy difference between the
highest occupied single particle orbital and the lowest unoccupied
orbital (known as HOMO-LUMO gap in molecular physics).  The neutrons
show always the smaller gaps and these lowest 1-particle-1-hole
($1ph$) transitions take the lead in the composition of the lowest
$2^+$ state. Correspondingly, both quantities share the same trends.
The spectral gap, in turn, is related to the effective mass $m^*/m$ of
the forces. We see that also in table \ref{tab:e2p} where low $m^*/m$
correlate to large gaps and vice versa.  But the step down to the
rather low spectral gap for SkO is much larger than the step up in
effective mass. Here we see also an interference from the very strong
isovector spin-orbit force of SkO, another important contributor to
shell effects.
The table \ref{tab:e2p}, furthermore, demonstrate the effect of the
residual interaction in that the $E(2^+)$ are generally 1.5 MeV below
the lowest $1ph$ energy. This allows to postulate a simple criterion
for the selection of forces: the lowest spectral gap in $^{132}$Sn
(and other doubly magic nuclei) should stay safely above the
experimental $E(2^+)$, which is 4.04 MeV for $^{132}$Sn. The force SkO
clearly fails in that respect. The reason is that SkO was fitted to
match the two-nucleon shell gaps at doubly magic $^{208}$Pb already at
the level of pure mean-field calculations \cite{Rei99b}.
Meanwhile, it has been shown that collective correlations reduce the
two-nucleon shell gaps by 1-2 MeV \cite{Fle04a}. The fitting strategy
of SkO thus squeezes the spectral gap too much with the obvious
consequence that the collective spectra are spoiled throughout. 
This mismatch is thus a strong hint on the inner coherence of Skyrme
forces connecting the various observables.

Far away from the magic $N=82$, one expects that the pairing gap
dominates the $E(2^+)$ energy. The pairing force was tuned in the same
way to the odd-even staggering in Sn isotopes and $N=82$ isotones.
Thus the pairing gap is about the same for all four Skyrme forces in
the well pairing region ($N<80$). We see indeed comparable energies
for the three forces, SkM$^*$, SLy6, and SkI3, which also hit very
nicely the experimental values. The force SkO, however, produces
systematically larger energies out there. This shows that shell
effects (here probably from the spin-orbit force) have also some
influence.  Different relations are seen in the other pairing regime
for the neutron rich, exotic nuclei above $N=82$ where SkM$^*$ shows
always the largest energies. This is at the same time a region of weak
binding. This causes a strong interplay of shell effects and pairing
which are not easily disentangled. 

Large differences are seen in the immediate vicinity of $N=82$. All
forces, expect SkO, reproduce nicely the sudden step down from $N=82$
and the asymmetry around $N=82$, namely the fact that $N=84$ has lower
$E(2^+)$ than $N=80$. But the results differ in the trends for
$N=78$. The case SkI3 follows nicely the smooth experimental trend
while SkM$^*$ and SLy6 show a spike.  That is compensated at $N=70$
where now SkI3 has a spike. In all these cases, we found that a larger
$E(2^+)$ is related to a somewhat lower neutron level density at the
Fermi surface.

A much more critical test than excitation energies are the associated
transition probabilities, the $B(E2)$ values. One is usually happy to
describe them within a factor of 2 or so, and often the concept of
effective charges is introduced to achieve a fine tuning
\cite{Ham96a}. The lower panel of figure \ref{pic:all-e2p} shows the
$B(E2)\!\!\uparrow$ values for the transitions. For $N\leq 82$, they 
are similar for all four forces in spite of the sometimes very different 
energies.  But they all differ from the experimental data by about a 
factor of two. The positive aspect is that the theoretical results 
come so close at all in view of the fact that the transition strengths 
are always much more demanding.
The remaining mismatch can have various origins: 
1) We use simply a quadrupole constraint to generate the collective
path instead of the variationally optimized ATDHF prescription
\cite{Bar78a,Goe78a,Kle91a}.
2) We use the raw quadrupole expectation value rather than the
fully mapped collective image (see appendix \ref{sec:obser}).
3) The effective energy functional is not fully suited to compute
transition moments and effective charges had to be added for
a correct description \cite{Ham96a}. 
Which one of these approximations is most responsible, has yet to be
explored. Anyway, the results are not untypically bad because almost
all microscopic approaches have a hard time with an exact reproduction
of transition moments.

The $B(E2)\!\!\uparrow$ in the region around the doubly magic
$^{132}$Sn surprisingly shows basically no differences between the
forces, just in a region where the energies differ most. A very
interesting point is the magic $^{132}$Sn. Naive models predict a
dramatic drop in the $B(E2)$ at the magic point. A model study taking
care of the residual interaction and cross-talk between the neutron
and proton quadrupole vibrations predicts that the $B(E2)$ should,
quite oppositely, have a peak at $^{132}$Sn \cite{Teras02}. Our
calculations confirm these estimates at a qualitative level, namely to
the extent that we also do not find any deep dip in the $B(E2)$. In
our cases, the residual interaction was obviously not large enough to
turn that into a peak. But these are quantitative details of the
employed forces.

The largest differences between forces for the $B(E2)\!\!\uparrow$
values are seen in the deep exotic regime $A>132$. We are sure that
information about $B(E2)$ in that region would be valuable. But before
one can exploit that, one has to understand (and possibly remove) the
systematic overestimation still seen on the low-$A$ side.

\subsubsection{Variation of pairing recipes}

\begin{figure}
\unitlength1mm
\centerline{\includegraphics[width=85mm]{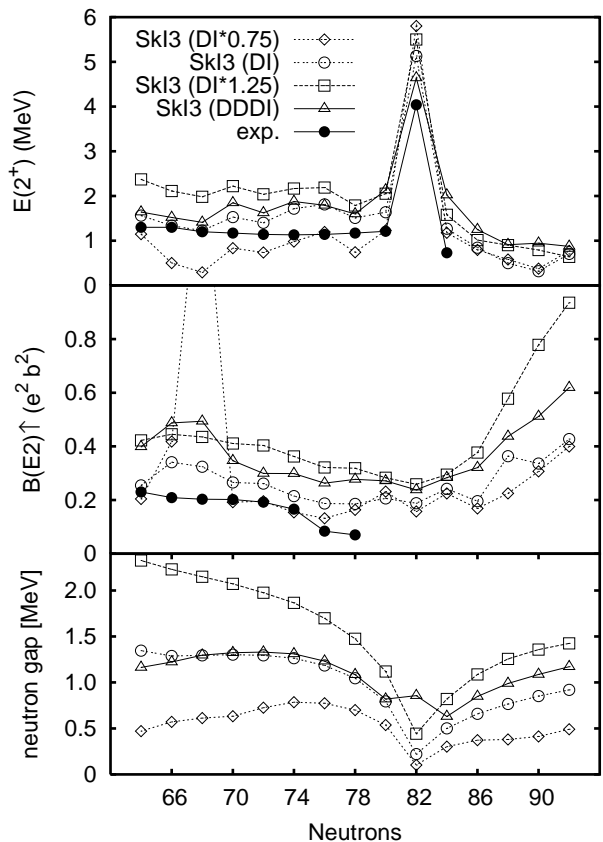}}
\caption[zu]{\label{pic:pair-e2p}
  Energies $E(2^+)$, $B(E2)\!\!\uparrow$ values
  (= $| \langle0^+|Q_{2M}|2^+\rangle |^2$), and
  average neutron-pairing gaps at spherical shape
  along the chain of 
  Sn isotopes calculated with SkI3 and three different
  pairing recipes: standard DI (circles), DDDI (boxes), and
  DI with 25\% enhanced strength.
  The experimental results are taken from \cite{nudat}.
}
\end{figure}
Figure \ref{pic:pair-e2p} shows the collective spectra along the Sn
chain for SkI3 computed with different pairing
prescriptions. We have added in the lowest panel
some information about the internal pairing structure, namely the
average neutron-pairing gaps 
$\bar\Delta=
  \sum_\alpha u_\alpha v_\alpha \Delta_{\alpha\alpha}\big/
  \sum_\alpha u_\alpha v_\alpha 
$
which are deduced from spectral properties of the given nucleus 
at th spherical shape (usually the minimum n the PES) and
which, nonetheless, provide a simple measure for the pairing gap
deduced from odd-even staggering \cite{Ben00c}.
The deliberately changed pairing strengths (boxes versus circles for
enhanced pairing and rhombus versus circle for reduced pairing) has an
obvious effect. The pairing gap is increased or reduced and
subsequently, the $E(2^+)$ energies change in the same direction.  The
effect is most pronounced in the regions sufficiently far off $N=82$
where we expect a dominance of pairing in the collective spectra.  The
relative changes in excitation energies and pairing gaps are much
larger than the change in pairing strength. Moreover, the excitation
energies behave generally similar to the pairing gaps.  This
demonstrates that the low-energy spectra in soft vibrators provide
valuable information about the pairing strengths.  On is tempted to
use that for an immediate tuning of the strengths.  We run however, in
some conflict, because $E(2^+)$ energies and $B(E2)\!\!\uparrow$ do no
coincide at the same strength.  Moreover, one has to keep in mind that
the information from lying states is still mixed with effects of the
mean field.  This is a general feature of nuclear structure, even for
the odd-even staggering which is usually taken as benchmark for
pairing properties \cite{Dob01a}.  The interference of shell effects
can be seen here particularly well from some irregularity at
$N=68$. The reduced pairing produces for $N=68$ a strongly deformed
ground state which results in a sudden drop of the $E(2^+)$ energy
accompanied by a strong peak in $B(E2)\!\!\uparrow$. The average gap
shows no dramatic reaction because it remains related to the now
irrelevant spherical shape.
The DDDI pairing stays in most cases more or less close to the results
of DI pairing which is somewhat expected because it is tuned to the
same average pairing gap. However, DDDI pairing reacts differently to
shell effects as can be seen in the vicinity of the shell closure
$N=82$ and for the weak shell closure which SkI3 produces at
$N=72$. The DI pairing seems to comply better with data. But that
holds in connection with the particular shell structure of SkI3. Much
more systematic investigations with varied forces and in other region
of the nuclear chart are necessary before drawing any conclusion like
that.
The $B(E2)\!\!\uparrow$ values shown in the lower panel of figure
\ref{pic:pair-e2p} show generally smooth trends, except for $N\geq 86$
where an increasing trend sets on which is related to the increasing
softness of these neutron rich isotopes.  The sensitivity to varying
pairing recipes is similar to what we have seen when varying the
forces: They vary little for $N\leq 82$ and more significant
differences appear in the far exotic regime $N> 82$.

\begin{figure}
\unitlength1mm
\centerline{\includegraphics[width=85mm]{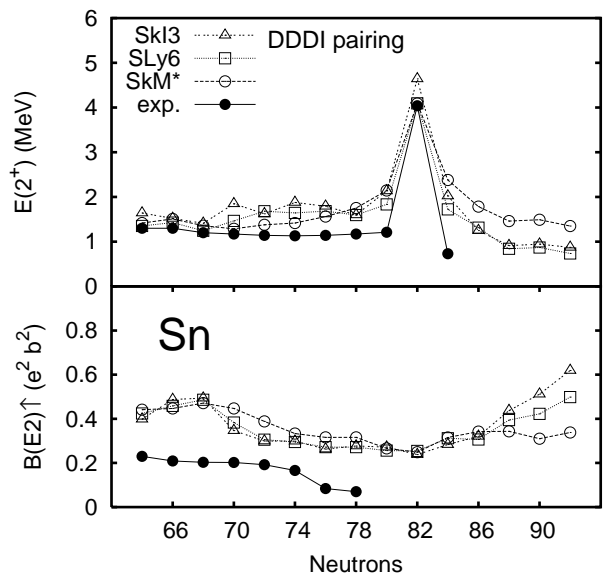}}
\caption[zu]{\label{pic:all-e2p_DDDI}
 As figure \ref{pic:all-e2p}, but now with DDDI pairing.
}
\end{figure}
For completeness, it is worth to look at the performance of DDDI
pairing also for the other Skyrme forces in the survey. That is shown
in figure \ref{pic:all-e2p_DDDI}. It has to be compared with figure
\ref{pic:all-e2p}. Similarities and differences are about the same for
all shown forces. The average excitations in the well pairing regime
are comparable. The small fluctuations about the average trends appear
for DDDI at different places than for DI pairing. The most pronounced
difference to DI is seen for the $E(2^+)$ energy next to the magic
neutron number, i.e. for $N=80$ and 84. DI pairing reproduces the
steep experimental drop while the DDDI results make all a somewhat
less dramatic step.
The $B(E2)\!\!\uparrow$ values shown in the lower panel are very
similar to those from DI pairing. They seem to be here the more robust
signal. We conclude from this results that the $B(E2)\!\!\uparrow$ are
insensitive to pairing while they are the much more sensitive
observable in other respects, e.g. in its dependence on the force.

A final comment on the $B(E2)\!\!\uparrow$ values: Good vibrators and
well-developed rotators are distinguished by the fact that the lowest
$2^+$ state exhausts the quadrupole sum rule in collective space. The
test for this feature is the comparison of the variance (= quadrupole
sum rule) with the $B(E2)$ value
\begin{equation}
  \langle0^+|Q_{20}^2|0^+\rangle
  =
  \sum_n\left|\langle 2^+_n|Q_{20}^{\mbox{}}|0^+\rangle\right|^2
  \stackrel{?}{=}
  \left|\langle 2^+_1|Q_{20}^{\mbox{}}|0^+\rangle\right|^2
\label{eq:estBE2}
\end{equation}
where $\langle...\rangle$ means the average in collective space, $0^+$
the ground state, $2^+_n$ the spectrum of $2^+$ states, and $2^+_1$
the lowest $2^+$ state. We have checked that and found that there is
generally good exhaustion of the variance by the lowest $2^+$.  The
collective potential ${\cal V}(\alpha_{20})$ of all these isotopes
excludes rotors, so that we can conclude that they are good vibrators.

\subsubsection{Effect particle-number correction}
\label{sec:pncorr}

The collective Schr\"odinger equation contains the
particle-number correction with $\hat{N}_{\rm coll}$ as discussed in
section \ref{sec:partnumb}.  It is interesting to
check the impact on collective properties.
\begin{figure}[h!]
\unitlength1mm
\includegraphics[width=90mm]{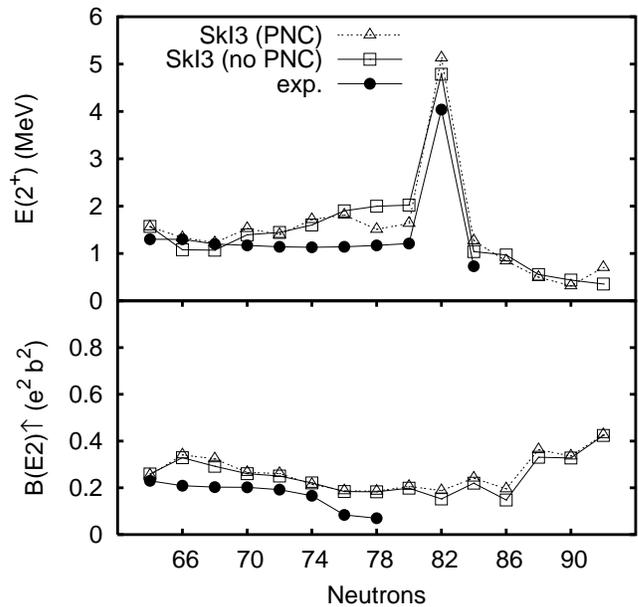}
\caption[zu]{\label{pic:Sn_num_cor}
  Energies $E(2^+)$ and $B(E2)\!\!\uparrow$ values
  (= $| \langle0^+|Q_{2M}|2^+\rangle |^2$)
  along the chain of Sn isotopes computed with SkI3 and DI pairing,
  once with (solid line, full circles) and once without 
  (dashed line, open circle) particle number restoration
  as outlined in section  \ref{sec:partnumb}.
  The Experimental results (dotted lines, full squares)
  are taken from \cite{nudat}.}
\end{figure}
This is done in figure \ref{pic:Sn_num_cor}. There is minimal
difference for the $B(E2)\!\!\uparrow$ values.  The main effect is seen
for energies in the region of weak pairing, i.e. at and around shell
closure. In fact, the particle-number corrected treatment seems to me
a bit more sensitive to shell structure as can be seen from the fact
that calculations without the correction shows generally smoother
trends. But this statement should be taken with a grain of salt. The
differences are anyway not very dramatic in view of the effects we see
when comparing forces and pairing recipes.

\subsection{Results for the isotopes of Cd and Te}

%
\begin{figure}[h!]
\unitlength1mm
\includegraphics[width=80mm]{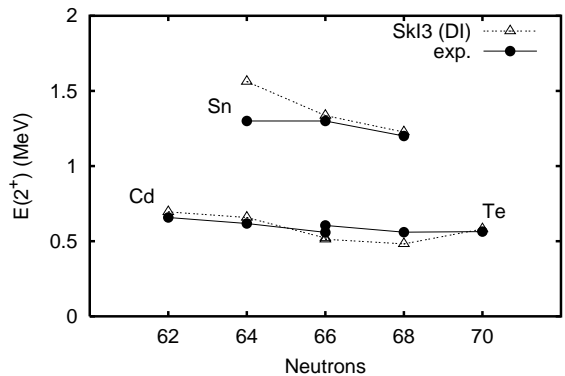}
\caption{\label{pic:E2p-syst}
            Systematics of the energies of the first excited \tp states
            calculated with the microscopic Bohr-Hamiltonian (\ref{eq:hcoll})
            using the interaction SkI3 for the nearest even-even 
            neighbors of $^{116}$Sn.
            The Experimental results are taken from \cite{nudat}.
}
\end{figure}
The Sn isotopes have a magic proton number $Z\!=\!50$.  It is
interesting to have a look at its even neighbors, Cd with $Z\!=\!48$
and Te with $Z\!=\!52$. As a first impression, we show in figure
\ref{pic:E2p-syst} a direct comparison for a few selected
isotones.
The effect is obvious. The step from the magic proton number to the
non-magic ones reduces once more the $E(2^+)$ energies by a
substantial factor, fully in agreement with the experimental findings.
The quadrupole mode in Cd and Te is much softer than in Sn where the
magic proton number enhances the rigidity of the whole mode due to a
strong residual proton-neutron force \cite{wood92}.

\begin{figure}
\unitlength1mm
\includegraphics[width=85mm]{{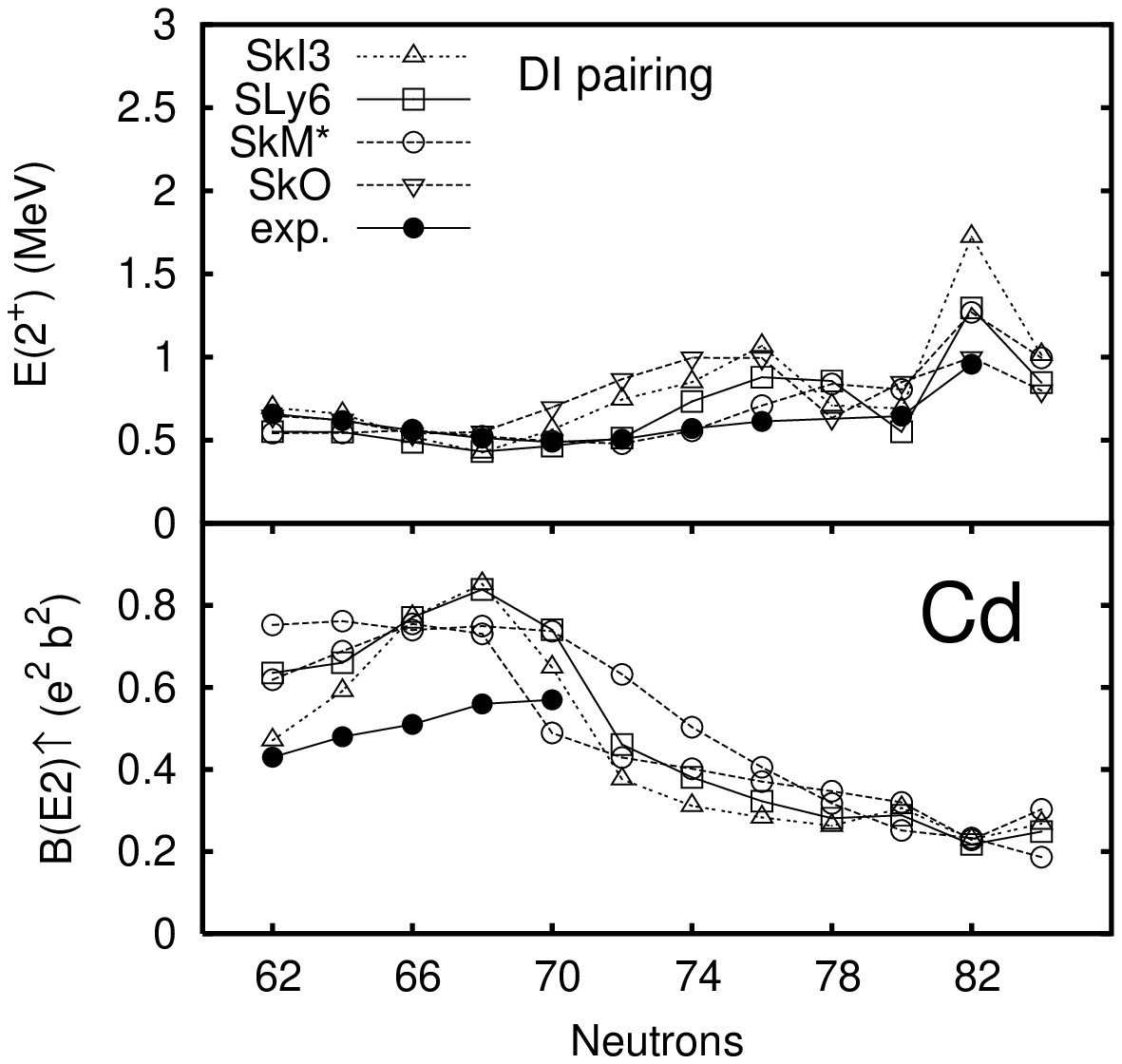}}
\\
\includegraphics[width=85mm]{{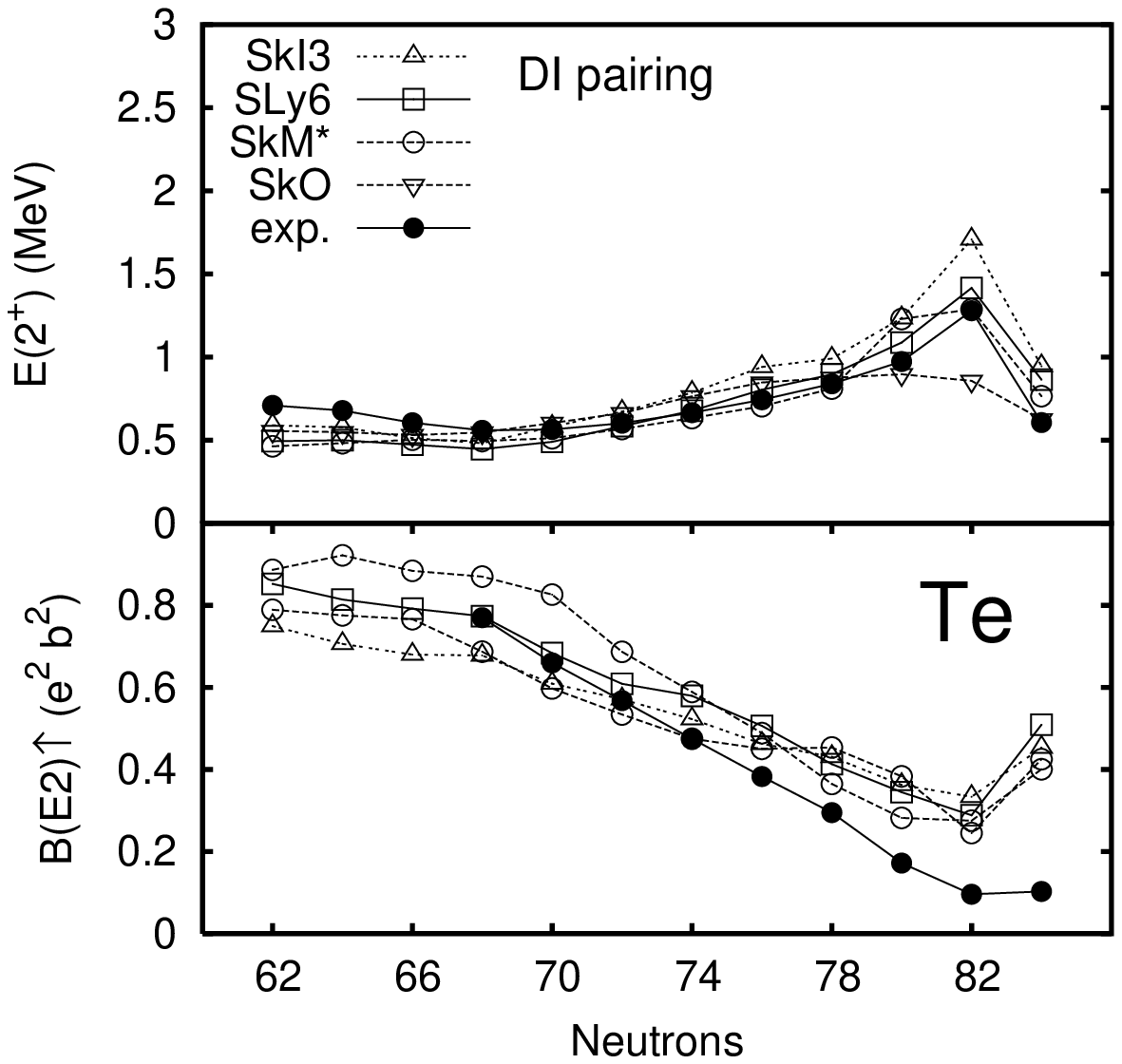}}
\caption[zu]{\label{pic:Cd_Te_all} 
  Energies $E(2^+)$ and $B(E2)\!\!\uparrow$ values
  (= $| \langle0^+|Q_{2M}|2^+\rangle |^2$)
  along the chain of 
  Cd isotopes (upper panels) and
  Te isotopes (lower panels) 
  calculated with different Skyrme forces as indicated.
  The experimental results are taken from \cite{nudat}.
}
\end{figure}
A summary of results for the Cd and Te isotopes using the four
different Skyrme interactions SkI3, SLy6, SkM*, and SkO is displayed
in Fig. \ref{pic:Cd_Te_all}. For Te, the $E(2^+)$ from different
forces are very close to each other and to the experimental data up to
$N\!=\!76$ and show again larger differences near the shell closure
$N\!=\!82$. The case for Cd is similar showing, however, an earlier
onset of differences. It is noteworthy that the results for SkO reside
well amongst the other forces, although it behaved dramatically
different for the Sn isotopes. This is related to the fact that shell
effects are somewhat suppressed in Cd and Te because these have
non-magic proton number.
Figure \ref{pic:Cd_Te_all} shows also as complementing information the
$B(E2)$ values for the Cd and Te isotopes.  The differences between
the theoretical and the experimental values are for Cd in the same
order of magnitude like in the case for the Sn chain. They tend to be
much less for the Te isotopes for $N\!\leq\!78$ where deviations stay
below 20\%. This looks like a remarkable agreement. But both,
agreement for Te and disagreement for Cd and Sn have yet to be
understood in detail.

\begin{figure}
\unitlength1mm
\includegraphics[width=85mm]{{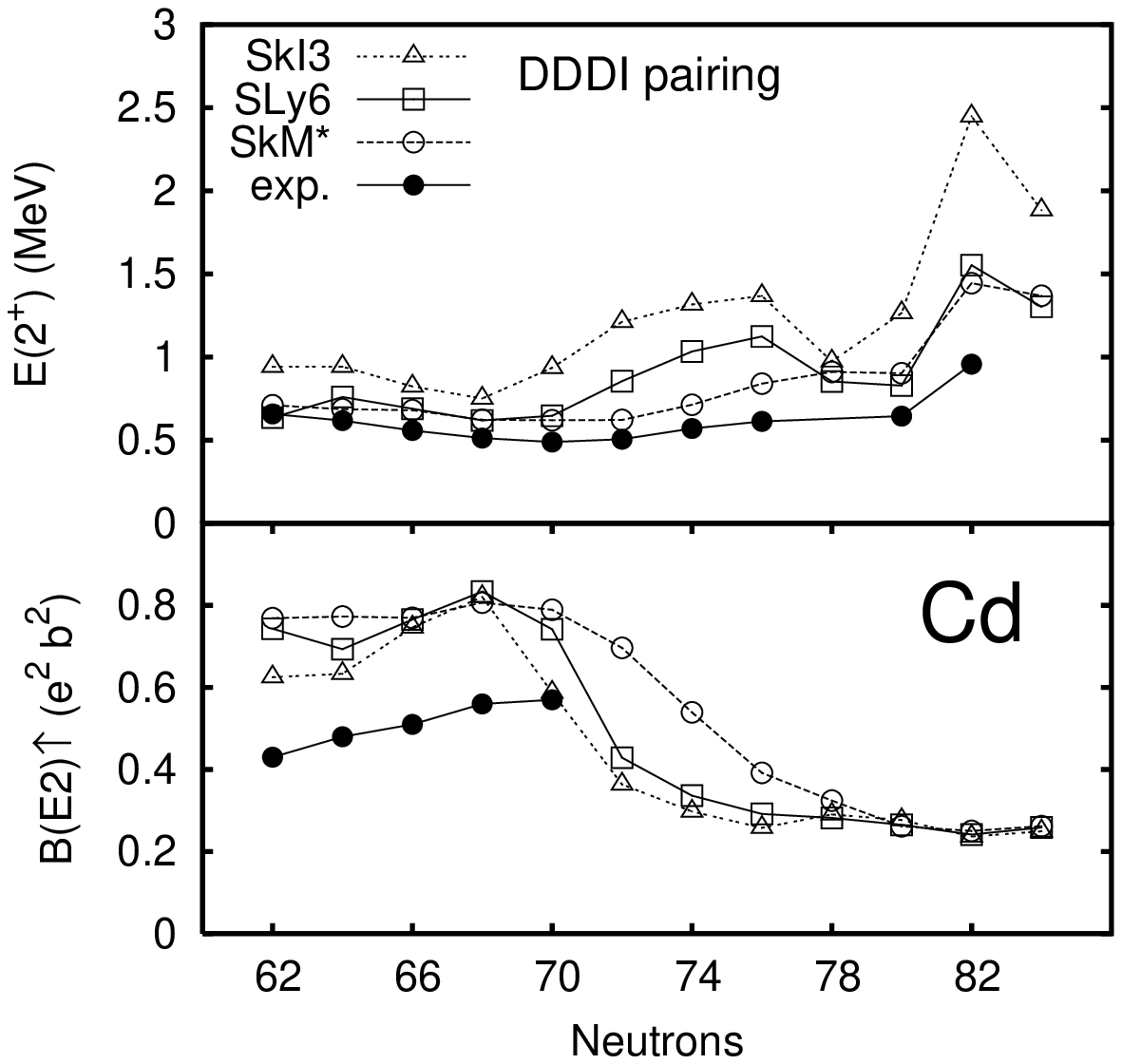}}
\\
\includegraphics[width=85mm]{{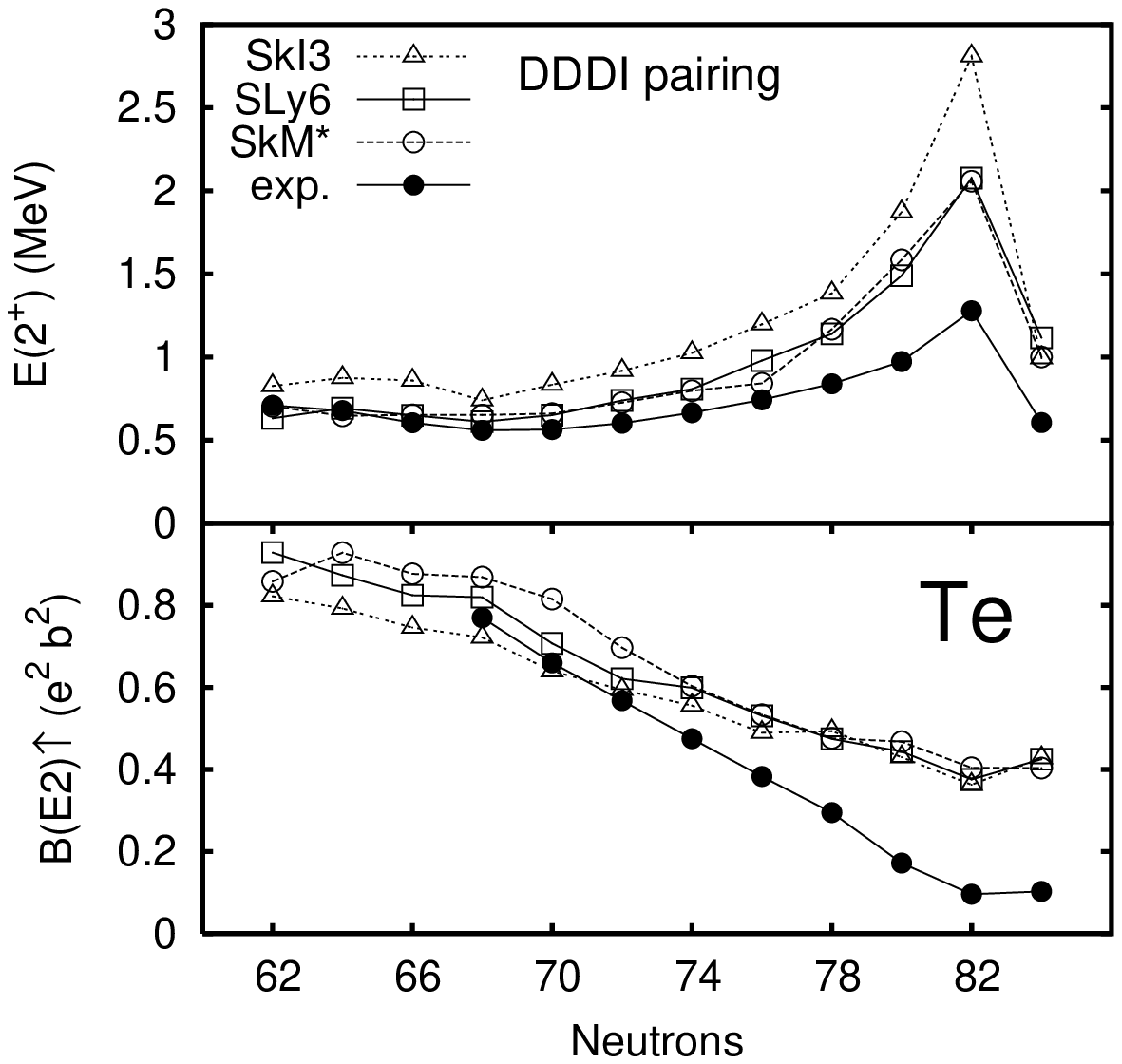}}
\caption[zu]{\label{pic:Cd_Te_all_DDDI} 
As figure \ref{pic:Cd_Te_all}, but now for DDDI pairing.
}
\end{figure}
We have seen in the Sn chain that the step from DI to DDDI pairing
makes most differences next to the neutron shell closure. One has to
suspect that a similar feature appears next to the proton shell
closure.  The Sn chain reside at $N=50$ which is a closed proton
shell. Thus the neighbor chains for Cd an Te are in the most
sensitive regime and we expect visible differences. The results on the
low lying $2^+$ states along Cd and Te are shown in figure
\ref{pic:Cd_Te_all_DDDI}. We see indeed that the $E(2^+)$ energies are
larger with DDDI, particularly near the neutron shell closure at
$N=82$. The differences become again negligible far out in the well
pairing regime. And, as for Sn, the $B(E2)$ values are totally
insensitive. It is also clear that the DI results, here and in the Sn
chain, are closer to the data. The same effect was already seen for the 
neutron channel in figure \ref{pic:pair-e2p}. The step of the $E(2^+)$ when 
moving away from a magic number is softer for DDDI than for DI. This
produces somewhat to high $E(2^+)$ for $N=80$ and $84$ in figure 
\ref{pic:pair-e2p} and here for $Z=48$ and $52$ in figure 
\ref{pic:Cd_Te_all} vs. figure \ref{pic:Cd_Te_all_DDDI}. We just 
remark this observation.  It is to early to draw far reaching conclusions 
on the validity of DI versus DDDI. The difference is seen in the worst 
case, namely a nucleus in the weak pairing regime where we are not yet 
sure that the present pairing treatment (BCS plus Lipkin-Nogami) is 
fully appropriate.

\subsection{The isotope shifts for the Sn isotopes}
\label{ss:iso-sn}

The ground state solution of the collective Schr\"odinger equation
provides the collective ground state correlations. One can compute the
correlation effect on any one-body observable with the help of the
collective map as outlined in section \ref{sec:obser}.
\begin{figure}[h!]
\unitlength1mm
\includegraphics[width=85mm]{{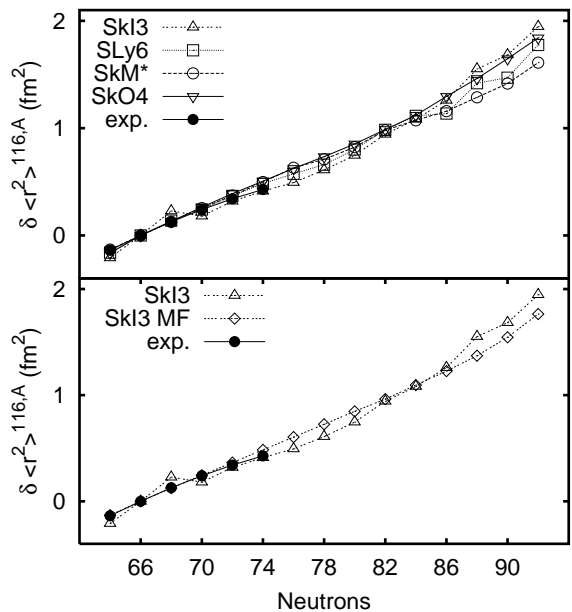}}
\caption[zu]{\label{pic:sn-iso}
Isotope shifts of the charge r.m.s radii, 
$\delta\langle r^2_{\rm rms}\rangle^{66,N}$, relative to
$^{116}Sn$. Upper panel: comparison of results from 
the four forces SkI3, SLy6, SkM*, and SkO including
collective ground state correlations.
The expectation value $\langle r^2 \rangle$ is calculated here
according to \ref{sec:obser}.
Lower panel: Comparison of pure mean field result with those including
correlations for the force SkI3.
The experimental data is taken from \cite{eberz87}}
\end{figure}
Figure \ref{pic:sn-iso} shows results for the systematics of charge
r.m.s. radii drawn in terms of isotope shifts relative to $^{116}$Sn.
The r.m.s. radii shown in the upper panel here are taken from the
correlated ground state which includes the collective shape
fluctuations.  At first glance, all forces reproduce the experimental
trend very well where data are available. There remain small
but significant differences between the four forces in that range
amongst which SkI3 comes generally closest to the data.  The
similarity of the trends persists to isotopes with larger neutron
numbers. Substantial differences develop at the upper end for the
shown chain, i.e. for $N>86$. Not surprisingly, this is the regime of
exotic nuclei because the generally soft binding amplifies small
differences in shell structure. It is surprising, however, that these
differences develop so late. The regime of similarity reaches well
beyond the $N=82$ shell closure.  This is due to the smoothing
features of the shape fluctuations.
The lower panel demonstrates the effect of ground state correlations
for that observable. There are practically no visible effects as
compared to pure mean field calculations, as one could have expected
for such a chain of semi-magic nuclei \cite{Rei95a}. Correlation
effects become visible again at the upper edge of the chain where the
general softness of the deeply exotic nuclei does also allow for
larger shape fluctuations.

\section{Conclusion}

We have investigated the predictive power of nuclear effective forces
for describing low-lying collective states considering as test cases
the chain of Sn isotopes as well as its even neighbors Cd and Te.  As
a particular example of such an effective force we used the
Skyrme-Hartree-Fock scheme augmented by a short-range pairing force.
To that end, we used a representative sample of different Skyrme
forces as well as two different pairing models (volume pairing versus
density-dependent pairing). The spectra of collective quadrupole
vibrations were computed in a two step procedure: First, mean-field
calculations with quadrupole constraint and self-consistent cranking
where performed which provide the microscopic input for a collective
Hamiltonian in terms of potentials, masses and moments of
inertia. Second, the collective Schr\"odinger equation thus obtained
is solved in the space of the five quadrupole coordinates. Care has
been taken to subtract correctly the zero-point energies from spurious
collective fluctuations in the mean-field states and to respect the
topology of the quadrupole space. As a simplification, we use axially
symmetric mean-field calculations and interpolate triaxial properties
between prolate and oblate shapes. This is an acceptable approximation
for the nearly spherical soft vibrators considered in the present
survey. Pairing is treated at the level of BCS with particle number
correction in terms of the Lipkin-Nogami approximation. A final
fine-tuning of the average particle number is performed also for the
collective states.

We find three regimes: collective properties are dominated by the
pairing gap for $N\!\leq\!76$, they are dominated by the spectral gap
of the neutron level for $76\!\leq\!N\!\leq\!82$, and a sudden
transition to prolate ground state deformations emerges for
$N\!>\!82$. In the pairing dominated regime, the results for the $2^+$
excitation energies depend mostly on the pairing strength and only
weakly on the Skyrme forces (with exception of the force SkO which
behaves a bit strange in consequence of the constraint on two-nucleon
shell gas in the fit). As all pairing models used here were fitted to
the odd-even staggering in the Sn region, we find generally nice
agreement with experimental data in that mid shell region.
In the shell-gap dominated regime, on the other hand, a strong
sensitivity to the Skyrme force develops due to a strong relation to
the spectral gap which, in turn, depends sensitively on the effective
mass. These features persist into the regime above $N\!=\!82$. The
result for the transition probabilities (the $B(E2)$ values) show
larger deviations from the data (up to a factor of two). This is not
surprising because B(E2) values are generally more demanding to any
model.

To summarize, we have shown that Skyrme forces have, in principle, the
capability to describe low lying collective spectra. In practice, the
success depends on the actual parameterization used. Turning the
argument around, we find that a systematic investigation of collective
spectra delivers extremely useful information for the selection of
parameterizations and the development of improved effective
forces. This calls for more systematic investigations.

\bigskip

\noindent
Acknowledgment:\\
We thank M. Bender, T. B\"urvenich, and T. Cornelius for many
clarifying and inspiring discussions.
This work was supported
in part by the Bundesministerium f\"ur Bildung und Forschung (BMBF),
Project Nos.\ 06 ER 808 and 06 ER 124.

\appendix
\section{Microscopic computation of collective operators}
\label{app:method}

\subsection{The deformed mean field}

Microscopic basis is a self-consistent mean-field according to SHF with DI or
DDDI pairing (for details see \cite{Ben03aR}).  The SHF-BCS equations describe
the nuclear state in terms of a set of single particle states $\varphi_n$ with
associated BCS occupation amplitudes $v_n$. These together compose the BCS
state
$|\Phi\rangle=\prod_n\left(u_n+v_n\hat{a}_n^+\hat{a}_{-n}^+\right)
|\mbox{vac}\rangle$
where $u_n^{\mbox{}}=\sqrt{1-v_n^2}$.  As a synonym for its content we denote
it by $|\Phi\rangle\equiv\{\varphi_n,v_n\}$.  In practice, we go somewhat
beyond the BCS scheme by using the Lipkin-Nogami (LN) approximation for
particle number projection \cite{Rei96a}.
The mean-field equations can be summarized as
\begin{subequations}
\label{eq:mfeqs}
\begin{equation}
  \left(\hat{h}-
   \sum_\nu(\epsilon_{{\rm F},\nu}\hat{N}_\nu-\epsilon_{2,\nu}\hat{N}_\nu^2)
   -\lambda\hat{Q}_{20}\right)
  |\Phi_{\alpha_{20}}\rangle
  =
  {\cal E}|\Phi_{\alpha_{20}}\rangle
\label{eq:mfeq}
\end{equation}
\begin{eqnarray}
  \hat{Q}_{20}
  &=&
  r^2Y_{20}f_{\rm cut}({\bf r})
  \;,
\label{eq:quad}\\
  \alpha_{2m}
  &=&
  \frac{4\pi}{5}
  \frac{\langle\Phi_{\alpha_{20}}|r^2Y_{2m}|\Phi_{\alpha_{20}}\rangle}
       {Ar^2}
  \;,
\label{eq:label}
\end{eqnarray}
The $\hat{h}$ is a two-quasiparticle operator which itself depends on the
state on which is acts. The actual form is obtained by a functional derivative
of the given energy functional.  
The $\hat{N}_\nu$ is the operator of proton- or neutron-number.
The Fermi energy $\epsilon_{{\rm F},\nu}$ is chosen
to tune the correct particle number in the average, i.e.
\begin{equation}
  \epsilon_{{\rm F},\nu}
  \quad\longleftrightarrow\quad
  \langle\Phi|\hat{N}_\nu|\Phi\rangle=N_\nu
  \;.
\label{eq:matchN}
\end{equation}
\end{subequations}
For simplicity we write in the following one particle-number term as
representative of both.
The term $\propto\hat{N}_ \nu^2$ accounts for the approximate
particle-number projection and its parameter $\epsilon_{2,\nu}$ is
given according to the LN recipe taking properly into account the
feedback from the mean field to the variances \cite{Rei96a}. The LN
scheme performs also an approximate variation after projection. This
yields a finite pairing gap under any conditions, even at shell
closures. And this is the feature what we need to have a smooth
evolution of the gap along the collective deformation path. Pure BCS
can lead to discontinuities which lead to discontinuities in the
collective Hamiltonian.

The stationary mean field equations without constraint provide only a
few well isolated states, the ground state and perhaps some isomers.
In order to describe motion, one needs to consider a time-dependent
mean-field theory, as e.g. time-dependent Hartree-Fock (TDHF). Large
amplitude collective motion is related to low energy excitations, thus
slow motion. This justifies the adiabatic limit known as adiabatic
TDHF (ATDHF) \cite{Bar78a,Goe78a}. It yields at the end, an
optimized constraint $\hat{Q}$ and subsequent collective path
$\{|\Phi_q\rangle\}$ where $q$ stands for continuous series of
deformations.  The fully self-consistent optimization of the path is
very cumbersome. It is a widely used approximation to use a simple
quadrupole constraint as in (\ref{eq:quad}). The anomalies at large
distance are avoided by a cut-off function $f_{\rm cut}$ for which we
use a Woods-Saxon shape \cite{Rut95a}.
The states are labeled with the dimensionless quadrupole moment
(\ref{eq:label}) which is rescaled with the total particle number $A$
and the r.m.s. radius $r$.  The index $m$ can run over $-2,-1,0,1,$
and 2. The path will be computed only along axially symmetric shapes
corresponding to $m=0$.

The numerical solution is done by standard methods. Wavefunctions and fields
are represented on an axially symmetric grid in coordinate space. An
accelerated gradient method is used to iterate the single particle states
$\varphi_n$ into their stable solution \cite{Rei82a} while the BCS+LN
equations for $v_n$ are solved in each iteration step.  An extra iterative
loop is included to maintain a wanted value of $\alpha_{20}$
\cite{Cus85a}. 

Knowing the path, yields immediately the raw collective potential as
\begin{equation}
  {\cal V}(\alpha_{20})
  =
  E_{\rm SHF}\left(|\Phi_{\alpha_{20}}\rangle\right)
\label{eq:rawpot}
\end{equation}
where $E_{\rm SHF}$ is the total SHF+BCS+LN energy for the given mean
field state $|\Phi_{\alpha_{20}}\rangle$ including a c.m. correction
as appropriate for the given force \cite{Ben03aR}. The actual
computations exploit explicit expressions in terms of the
single-particle states $\varphi_n$ and their occupation amplitudes
$v_n$.

\subsection{Computation of masses}

The collective path allows to define a collective momentum operator as
the generator of deformation
$
  \hat{P}_\alpha|\Phi_{\alpha_{20}}\rangle
  =
  {\rm i}\partial_{\alpha_{20}}|\Phi_{\alpha_{20}}\rangle
  \;.
$
At the same time, $\hat{P}_\alpha$ can also be interpreted as the momentum
operator associated with collective dynamics. The collective path
$\{|\Phi_{\alpha_{20}}\rangle\}$ is complemented by the dynamical response of
the system by adding a dynamical constraint 
$-\mu\hat{P_\alpha}$ to the mean-field equations
yielding eventually a dynamical collective path.
The adiabatic approximation allows to handle the 
dynamical part in the linear regime, i.e.
\begin{equation}
  |\Phi_{\alpha_{20}p_\alpha}\rangle
  \approx
  \left(1+{\rm i} p_\alpha\hat{Q}_\alpha^{\rm(dyn)}\right)
  |\Phi_{\alpha_{20}}\rangle
  \;.
\end{equation}
The solution of the linear response equation thus obtained provides
the dynamical response generator $\hat{Q}_\alpha^{\rm(dyn)}$. Note
that the whole energy functional is involved in the response. This is
called self-consistent, or ATDHF, cranking, see e.g.
\cite{Goe78a,Dob81a}. The inverse collective mass for quadrupole
motion is then obtained in straightforward manner as
\begin{eqnarray}
  B
  &=&
  \frac{1}{2}
  \frac{\partial^2 E_{\rm SHF}\left(|\Phi_{\alpha_{20}p_\alpha}\rangle\right)}
       {\partial{p_\alpha}^2}
  \Big|_{p_\alpha=0}
\nonumber\\
  &\equiv&
  \frac{1}{2}
  \langle\Phi_{\alpha_{20}}|
   [\hat{Q}_\alpha^{\rm(dyn)},[\hat{H},\hat{Q}_\alpha^{\rm(dyn)}]]
  |\Phi_{\alpha_{20}}\rangle
  \;.
\label{eq:collB}
\end{eqnarray}
The second form with the double commutator is not strictly applicable
in connection with energy functionals (where the full $\hat{H}$ is not
given). It serves here only as a notational abbreviation to establish
contact with standard formulae for cranking masses.

The same procedure is applied to the dynamical response to
rotations. The collective momentum is here already known as it is the
angular momentum, e.g $\hat{J}_x$.  Solving the equations for the
corresponding dynamical response yields the momentum of inertia as
\begin{eqnarray}
  \frac{1}{2\Theta_{xx}}
  &=&
  \frac{1}{2}
  \frac{\partial^2 E_{\rm SHF}\left(|\Phi_{\alpha_{20}\omega}\rangle\right)}
       {\partial\omega^2}
  \Big|_{\omega=0}
\nonumber\\
  &\equiv&
  \frac{1}{2}
  \langle\Phi_{\alpha_{20}}|
   [\hat{Q}_J^{\rm(dyn)},[\hat{H},\hat{Q}_J^{\rm(dyn)}]]
  |\Phi_{\alpha_{20}}\rangle
  \;,
\label{eq:collTh}
\end{eqnarray}
and similarly for $y$ and $z$.  The operator $\hat{Q}_J^{\rm(dyn)}$
carries the dynamical response in the same manner as
$\hat{Q}_\alpha^{\rm(dyn)}$ does that for the quadrupole motion. In
practice, we are considering only axially symmetric shapes
$\alpha_{20}$.  For then, we obtain 
$$
  \Theta_{xx}=\Theta_{yy}=\Theta
  \quad,\quad
  \Theta_{zz}=0
  \quad.
$$ For the case rotation, we simplify the response problem by
computing the response with the stationary mean field $\hat{h}$ only
(Inglis cranking). The approximation works very well for the
considered cases. The critical region of small deformations
\cite{Rei84b} does not contribute due to the topological switching
(\ref{eq:topswitch}).

For the full five-dimensional quantum corrections (see next subsection
\ref{sec:QC}), we also need to compute the inverse collective mass
$B_\gamma$ and width $\lambda_\gamma$ for vibrations in the
$\gamma$-direction.  We do that for the vicinity of axial shapes by
means of linear response. And we employ here the Inglis approximation
using only the mean-field Hamiltonian $\hat{h}$ in the response equations.
All together, we have then the necessary ingredients concerning
masses: the inverse masses $B$, $B_\gamma$ ($=B$), and the momentum of
inertia $\Theta(\alpha_{20})$.

\subsection{Quantum corrections}
\label{sec:QC}

The GCM ansatz for the collectively correlated state is written as a
coherent superposition over the path. However, the states of the path
correspond to wave packets in quadrupole space rather than to
eigenstates of $\alpha_{20}$ (and similarly for the dynamical
extensions in $p_\alpha$ and $\omega_{\rm crank}$). Thus they contain
spurious contributions from collective motion which contribute to any
expectation value. The strongest effects are found in the energy
expectation values which constitute the raw collective potential
(\ref{eq:rawpot}). These spurious contributions need to be
subtracted. That is what one calls the quantum corrections or
zero-point energies (ZPE) \cite{Rei87aR}.

The correction for spurious center-of-mass motion is already part of
the standard SHF scheme.  The most important for the collective
dynamics is the vibrational-rotational correction. These need to be
considered as one entity because vibrations and rotations are closely
connected pieces of the nuclear quadrupole topology.  The recipe for
purely axial vibration and rotation was given in
\cite{Rei78b,Rei87aR}. A recent model calculation has confirmed that
ansatz and proven that the correction provides also a very good
approximation to angular-momentum projection, again for heavy nuclei.
Here we want to account for the whole five-dimensional quadrupole
dynamics (see next section \ref{sec:topol}). Thus we are using the
properly generalized rotational-vibrational correction
\begin{subequations}
\label{eq:rotvibZPE}
\begin{eqnarray}
  E^{(\rm ZPE)}_{\rm quad}
  &=&
  \frac{\lambda_\beta}{4{\cal M}_\beta}
  +
  \frac{\partial^2_\beta{\cal V}}{4\lambda_\beta}
\\
  \lambda_\beta
  &=&
  2\langle\Phi_{\alpha_{20}}|\hat{P}_{\alpha}^2|\Phi_{\alpha_{20}}\rangle
\nonumber\\
  \hat{P}_\beta|\Phi_{\alpha_{20}}\rangle
  &=&
  i\partial_\beta|\Phi_{\alpha_{20}}\rangle
\nonumber\\
  E^{(\rm ZPE)}_{\rm triax}
  &=&
  \frac{\lambda_\gamma}{4{\cal M}_\gamma}
  +
  \frac{\partial^2_\gamma{\cal V}}{4\lambda_\gamma}
  \approx
  \frac{\lambda_\gamma}{4{\cal M}_\gamma}
\\
  E^{(\rm ZPE)}_{\rm rot}
  &=&
  \frac{\lambda_{\rm rot}}{4\Theta}
\\
  \lambda_{\rm rot}
  &=&
  2\langle\Phi_{\alpha_{20}}|\hat{J}_{x,y}^2|\Phi_{\alpha_{20}}\rangle
  \;.
\end{eqnarray}
\end{subequations}
where $\hat{J}_{x,y}$ is the angular momentum about the $x$- or $y$-axis.
The widths are the same for $x$ and $y$ because we evaluate everything
at axial symmetry, i.e. at the point $\gamma=0$. 
The total ZPE is decomposed as
\begin{subequations}
\label{eq:topswitch}
\begin{eqnarray}
  E^{\rm(ZPE)}_{\rm tot}(\alpha_{20})
  &=&
  \left(5-4g(\smfrac{\lambda_{\rm rot}}{4})\right)E^{\rm(ZPE)}_{quad}\\\nonumber
  &&+
  2g(\smfrac{\lambda_{\rm rot}}{4})E^{\rm(ZPE)}_{\rm rot}
  +
  2g(\smfrac{\lambda_{\rm rot}}{4})E^{\rm(ZPE)}_{triax}
  \;,
\\
  g(a)
  &=&
  \frac{\int_0^1 dx\, a(x^2-1)e^{a(x^2-1)}}{\int_0^1 dx\,e^{a(x^2-1)}}
  \;.
\end{eqnarray}
\end{subequations}
There is also a correction from spurious particle number
fluctuations. This is already taken into account in an approximate
manner by the Lipkin-Nogami scheme added on top of the BCS pairing.

All together, the quantum-corrected collective potential reads
\begin{equation}
  V(\alpha_{20})
  =
  {\cal V}(\alpha_{20})
  -
  E^{\rm(ZPE)}_{\rm tot}(\alpha_{20})
  \;.
\end{equation}
This is the quantity entering the collective Hamiltonian.  The masses
are associated with the collective kinetic energies which are already
of second order in the collective momenta. The quantum corrections on
masses would correspond to terms of fourth order and are neglected.

\subsection{Retuning the particle number}
\label{sec:partnumb}

All states along the collective path are tuned to have the same
average proton and neutron number. The energies are corrected by
approximate particle-number projection at the level of the LN
scheme. But the BCS states from which the collective path is composed
still carry these particle-number fluctuations. As a consequence, the
coherent superposition of the states along the path may change the
average particle number again. One needs to readjust the correct
average at the level of the collective dynamics \cite{Val00c}. To that
end, one builds the collective picture of the particle number operator
$\hat{N}-N$ in precisely the same manner as it was done for the
Hamiltonian. One obtains a particle-number potential, particle-number
masses for quadrupole as well as triaxial motion, and particle-number
contributions to the inertia. The expressions are the same as above
with $\hat{H}$ replaced by $\hat{N}$.
The collective image of $\hat{N}$ is added as a constraint in the
collective Schr\"odinger equation.

\subsection{The collective Schr\"odinger equation}
\label{sec:topol}

Axially symmetric quadrupole deformations are labeled by
$\alpha_{20}$. The full space of quadrupole deformations is explored
when considering all $\alpha_{2m}$ with $m\in\{-2,-1,0,1,2\}$.  This
is convenient for spherical vibrator nuclei as it implies
automatically the correct number of vibrational degrees of freedom.
It is, however, not well suited for deformed nuclei because rotations
look rather involved in that frame. It is customary to transform by
appropriate rotation into an intrinsic frame where
$a_{2\pm 1}=0$ and $a_{22}=a_{2\!-\!2}$. 
This defines three Euler angles {\boldmath$\vartheta$} as rotational
coordinates. The remaining two relevant deformation coordinates
$a_{20}$ and $a_{22}$ are expressed in terms of total deformation
$\beta$ and triaxiality $\gamma$ as $a_{20}=\beta\cos(\gamma)$ and
$a_{22}=\beta\sin(\gamma)/\sqrt{2}$ \cite{Rin80aB,Gre11B}.
Each triaxiality $\gamma$ which is an
integer multiple of $60^\circ$ corresponds to an axially symmetric
shape. The cases $\gamma=0^\circ, 120^\circ,$ and $240^\circ$
correspond to prolate axial deformations while $\gamma=60^\circ,
180^\circ,$ and $300^\circ$ are oblate.
Relevant information is contained in one $60^\circ$ sector of the
plane, e.g. in the segment $\gamma\in[0^\circ,60^\circ]$.
The other segments can be reconstructed by axis exchange of principle axes.
This symmetry under axis transformation has important consequences for
the representation of wavefunctions and potentials in the collective
Schr\"odinger equation: One has to obey mirror symmetry under
$\gamma\longrightarrow-\gamma$ and axis-rotation symmetry under
$\gamma\longrightarrow\gamma+120^\circ$.
The five-dimensional volume element $d^5a$ reads in the
$\beta$-$\gamma$-frame $d^5a=\beta^4
|\sin(3\gamma)|\,d\beta\,d\gamma\, d^3\theta$ where the $\theta$ are
the three Euler angles for the transformation from the laboratory
frame into the intrinsic frame.  In the following we will use the
notation $d^5\alpha$ as shorthand for the lengthy right hand side.

The collective Hamiltonian has the form of a Bohr-Hamiltonian
\cite{Boh52,Eis70aB} generalized to $\beta$-$\gamma$-dependent masses
(\ref{eq:hcolla}) \cite{pomorski,Liber99,Pro04a}.
\begin{widetext}
\begin{subequations}
\label{eq:hcoll}
\begin{eqnarray}\label{eq:hcolla}
    \hat{H}^{\rm(coll)} &=& -\frac{1}{\beta^4}\partial_\beta
    B(\beta,\gamma)\beta^4\partial_\beta -
    \frac{1}{\beta^2\,\sin3\gamma}\,\partial_\gamma
    B_\gamma(\beta,\gamma)\,\sin3\gamma\,\partial_\gamma +
    \sum_{k=1}^3 \frac{\hat{L}_k^{\prime
    2}}{2\Theta_k(\beta,\gamma)}+V(\beta,\gamma) \quad, \\[2.0ex]
  \label{eq:hcollb}
  X(\beta,\gamma)
  &=&
  \frac{X(\beta)\!+\!X(\!-\!\beta)}{2}
  +
  \frac{X(\beta)\!-\!X(\!-\!\beta)}{2}\cos(3\gamma)
  \qquad X \in \left\{B,V\right\} \quad,\\
  \label{eq:hcollc}
  \frac{1}{\Theta_k(\beta,\gamma)}
  &=&
  \frac{3}{4\sin^2\big(\gamma\!-\smfrac{2\pi}{3}k\big)} 
   \left[\frac{1}{2}
     \left(\smfrac{1}{\Theta(\beta)} + \smfrac{1}{\Theta(-\beta)}
     \right) 
     + 
     \left(
       \smfrac{1}{\Theta(\beta)}  - \smfrac{1}{\Theta(-\beta)}
     \right)\cos(\gamma\!-\smfrac{2\pi}{3}k)
   \right] 
   \qquad k \in \{1,2,3\}\label{eq:int_theta} 
\end{eqnarray}
\end{subequations}
\begin{subequations}\label{eq:psi-coll}
\begin{eqnarray}
  \bra{\,q\,}\,LM\rangle 
  &=& 
  \Psi^{LM}(\beta,\gamma,\vartheta_i) 
  = \sum_\nu \psi^{L}_{\nu}(\beta)\chi^{LM}_\nu\left(\gamma,\vartheta_i\right)
  \qquad \nu = 0,1,\dots
\\
  \left\{\chi^{00}_{\nu}\left(\gamma,\vartheta_i\right),\nu=0,1,...\right\}
  &=& 
  {\cal O} \left\{\cos(3\nu\gamma) D^{(0)}_{00}\right\} 
  = 
  \left\{\sqrt\frac{2\nu+1}{32\pi^2} P_\nu\left(\cos(3\gamma)\right)\right\} 
\\
  \left\{\chi^{2M}_{\nu}\left(\gamma,\vartheta_i\right),\nu=0,1,...\right\} 
  &=& 
  {\cal O} \left\{
    \cos(\lambda\gamma)\,D^{(2)}_{M0} -  (-1)^{\lambda\%3} 
    \sin(\lambda\gamma)\,\frac{D^{(2)}_{M,-2}+D^{(2)}_{M,2}}{\sqrt{2}}
  \right\}
\\
\nonumber
  &&
  \lambda=3\left[\frac{\nu}{2}\right]+\nu\%2+1
\end{eqnarray}
\end{subequations}
\end{widetext}
where ${\cal O}$ means ortho-normalization of the set,
$[...]$ the integer  part of a fraction, and $a\%b$ the modulo
of $a$ with respect to $b$.
The operator \(\hat{L}_k^\prime\) denotes the angular momentum in the
intrinsic frame. 
The deformed SHF calculations provide input along axially symmetric
deformations $a_{20}\stackrel{\small>}{\mbox{\tiny$<$}}0$. The
collective dynamics needs to be performed properly in all five
quadrupole degrees of freedom. No strong peaks or wells in the
$\gamma$ direction are to be expected for the nearly spherical or
weakly deformed soft vibrator nuclei which we will consider here. It
is thus an acceptable approach to interpolate the axial microscopic
results into the full $\beta$-$\gamma$ plane. While potential, inverse
masses and particle-number masses could be interpolated in a
straightforward manner by (\ref{eq:hcollb}), the moment of inertia has
three components. Respecting the collective symmetries we reconstruct
them from the axial information as (\ref{eq:hcollc}).  The deduction
from axial information implies that we neglect the $\beta$-$\gamma$
coupling in the kinetic energy, i.e. $B_{\beta\gamma}=0$.
The collective particle-number operator $\hat{H}^{\rm(coll)}$
is composed in the same form with the corresponding particle-number
masses, potentials and moments of inertia inserted.

The collective Schr\"odinger equations reads now
\begin{equation}\label{eq:solve}
  \left(
    \hat{H}^{\rm(coll)}-\epsilon^{\rm coll}_{\rm F}\hat{N}^{\rm(coll)}
  \right)\Psi
  =
  E\Psi
\end{equation}
where the correction of the Fermi-energy $\epsilon^{\rm coll}_{\rm F}$ is 
to be adjusted such that
$
  \int d^5a\,\Psi^+\hat{N}^{\rm(coll)}\Psi
  =
  N\;.
$

The dynamics is formulated in the whole \(\beta$-\(\gamma$-plane but
all necessary information is contained already in a segment of
$60^\circ$, as discussed in section \ref{sec:topol}. In order to meet
the inherent symmetry conditions \cite{Eis70aB}, the wavefunctions of
a $0^+$ and a $2^+$ state are expanded in a symmetrized base
(\ref{eq:psi-coll}) where the base mode $\nu=0$ determines the overall
$\beta$ dependence and the higher $\nu$ shape the profile in the
$\gamma$ direction. The Hamiltonian is very soft in $\gamma$ such that
few $\nu$-terms suffice for convergence (two or three, never more than
5). The $D^{(L)}_{M,K}(\mathbf{\vartheta})$ are the well known Wigner
$D$-Functions describing the rotation matrices for a state with
angular momentum $L$ \cite{edmonds}. It is noteworthy that the
structure of the rotational energy is that for the most general case
where the considered nuclei have no special symmetry. For that reason
the \tp state must be a sum over all possible z-projections of the
angular momentum which are $K=0,\pm 2$ in the intrinsic frame.

The remaining collective equation for the components
$\psi^{L}_{\nu}(\beta)$ is solved numerically with standard methods. The
wavefunctions and fields are represented on an equidistant grid in
$\beta$. Gradient iteration is used to find the few lowest eigenvalues
and states.

\subsection{Computation of observables}
\label{sec:obser}

The solution of the collective Schr\"odinger equation, as outlined in
the previous section, provides the energies directly. Expectation
values and transition moments of other observables $\hat{O}$ need yet
to be computed. The steps are, in principle, the same as done before
for the energy (= Hamiltonian $\hat{H}$ respectively) and for the
particle number $\hat{N}$.  One has first to determine the collective
image of the given observable $\hat{O}\longrightarrow O_{\rm coll}$
and computes then the expectation value of that image with the
collective wavefunctions \cite{Rei87aR}.  Actually, we do that for the
computation of the transition probabilities $0^+\longrightarrow 2^+$,
the proton B$(E2)$ values. The observable here is
$\hat{O}\equiv\hat{Q}_{20,{\rm prot}}$. In case of the isotope shifts
the observable is the collective mapping of the radius.
At present, we approximate the collective image by the raw expectation 
value only and neglect the kinetic corrections.

\bibliography{Sn_chain}

\end{document}